\documentclass[10pt,twoside,twocolumn,english,aps,manuscript,superscriptaddress,prbt,aps, preprint,manuscript,showpacs]{revtex4}
\usepackage[latin9]{inputenc}
\usepackage{color}
\usepackage{babel}
\usepackage{array}
\usepackage{textcomp}
\usepackage{amsbsy}
\usepackage{amstext}
\usepackage{amssymb}
\usepackage{graphicx}
\usepackage{esint}
\usepackage[unicode=true,pdfusetitle,
 bookmarks=true,bookmarksnumbered=false,bookmarksopen=false,
 breaklinks=false,pdfborder={0 0 1},backref=false,colorlinks=false]
 {hyperref}

\makeatletter

\DeclareFontEncoding{LGR}{}{}
\DeclareRobustCommand{\greektext}{%
  \fontencoding{LGR}\selectfont\def\encodingdefault{LGR}}
\DeclareRobustCommand{\textgreek}[1]{\leavevmode{\greektext #1}}
\ProvideTextCommand{\~}{LGR}[1]{\char126#1}

\newcommand{\lyxmathsym}[1]{\ifmmode\begingroup\def\b@ld{bold}
  \text{\ifx\math@version\b@ld\bfseries\fi#1}\endgroup\else#1\fi}

\providecommand{\tabularnewline}{\\}

\newcommand{\lyxdeleted}[3]{}

\@ifundefined{textcolor}{}
{%
 \definecolor{BLACK}{gray}{0}
 \definecolor{WHITE}{gray}{1}
 \definecolor{RED}{rgb}{1,0,0}
 \definecolor{GREEN}{rgb}{0,1,0}
 \definecolor{BLUE}{rgb}{0,0,1}
 \definecolor{CYAN}{cmyk}{1,0,0,0}
 \definecolor{MAGENTA}{cmyk}{0,1,0,0}
 \definecolor{YELLOW}{cmyk}{0,0,1,0}
}

\@ifundefined{date}{}{\date{}}
\makeatother

\begin{document}
\title{Magnetic properties of $S=5/2$ anisotropic triangular chain Bi$_{3}$FeMo$_{2}$O$_{12}$}
\author{K. Boya}
\affiliation{Department of Physics, Indian Institute of Technology Tirupati, Tirupati
517 506, India}
\author{K. Nam}
\affiliation{Department of Physics and Astronomy and Institute of Applied Physics,
Seoul National University, Seoul 151-747, Republic of Korea}
\author{A. K. Manna}
\affiliation{Department of Chemistry, Indian Institute of Technology Tirupati,
Tirupati 517 506, India}
\author{J. Kang}
\affiliation{Department of Physics and Astronomy and Institute of Applied Physics,
Seoul National University, Seoul 151-747, Republic of Korea}
\author{C. Lyi}
\affiliation{Department of Physics and Astronomy and Institute of Applied Physics,
Seoul National University, Seoul 151-747, Republic of Korea}
\author{A. Jain}
\affiliation{Solid State Physics Division, Bhabha Atomic Research Centre, Mumbai
400085, India}
\affiliation{Homi Bhabha National Institute, Anushaktinagar, Mumbai 400094, India}
\author{S. M. Yusuf}
\affiliation{Solid State Physics Division, Bhabha Atomic Research Centre, Mumbai
400085, India}
\affiliation{Homi Bhabha National Institute, Anushaktinagar, Mumbai 400094, India}
\author{P. Khuntia}
\affiliation{Department of Physics, Indian Institute of Technology Madras, Chennai
600 036, India}
\author{B. Sana}
\affiliation{Department of Physics, Indian Institute of Technology Madras, Chennai
600 036, India}
\author{V. Kumar}
\affiliation{Department of Physics, Indian Institute of Technology Bombay, Mumbai
400 076, India}
\author{A. V. Mahajan}
\affiliation{Department of Physics, Indian Institute of Technology Bombay, Mumbai
400 076, India}
\author{Deepak. R. Patil}
\affiliation{Department of Physics and Astronomy and Institute of Applied Physics,
Seoul National University, Seoul 151-747, Republic of Korea}
\author{Kee Hoon Kim}
\affiliation{Department of Physics and Astronomy and Institute of Applied Physics,
Seoul National University, Seoul 151-747, Republic of Korea}
\author{S. K. Panda}
\email{swarup.panda@bennett.edu.in}

\affiliation{Department of Physics, Bennett University, Greater Noida 201310, Uttar
Pradesh, India}
\author{B. Koteswararao}
\email{koteswararao@iittp.ac.in}

\affiliation{Department of Physics, Indian Institute of Technology Tirupati, Tirupati
517 506, India}
\date{{\today}}
\begin{abstract}
Competing magnetic interactions in low-dimensional quantum magnets
can lead to the exotic ground state with fractionalized excitations.
Herein, we present our results on an $S=5/2$ quasi-one-dimensional
spin system Bi$_{3}$FeMo$_{2}$O$_{12}$. The structure of Bi$_{3}$FeMo$_{2}$O$_{12}$
consists of very well separated, infinite zig-zag {\normalsize{}$S=5/2$}
spin chains. The observation of a broad maximum around $10$ K in
the magnetic susceptibility $\chi(T)$ suggesting the presence of
short-range spin correlations. $\chi(T)$ data do not fit to $S=5/2$
uniform spin chain model due to the presence of $2^{nd}$ nearest-neighbor
coupling ($J_{2}$) along with the $1^{st}$ nearest-neighbor coupling
($J_{1}$) of the zig-zag chain. The electronic structure calculations
infer that the value of $J_{1}$ is comparable with $J_{2}$ ($J_{2}/J_{1}\thickapprox1.1$)
with a negligible inter-chain interaction ($J'/J\thickapprox0.01$)
implying that Bi$_{3}$FeMo$_{2}$O$_{12}$ is a highly frustrated
triangular chain system. The absence of magnetic long-range ordering
down to $0.2$ K is seen in the heat capacity data, despite a relatively
large antiferromagnetic Curie-Weiss temperature $\theta_{CW}\thickapprox\lyxmathsym{\textendash}\,40$
K. The magnetic heat capacity follows nearly a linear behavior at
low temperatures indicating that the $S=5/2$ anisotropic triangular
chain exhibits the gapless excitations.
\end{abstract}
\maketitle

\section{Introduction}

Investigating the exotic magnetic properties of low-dimensional and
geometrically frustrated spin systems is one of the active research
fields in modern condensed matter physics \cite{A.Vasiliev NPJ 2018,A. Vasiliev handbook 2019,L. Balents Nature 2010,C. Broholm Science 2020}.
Mermin-Wagner theorem states that the system with dimensionality $d\leq2$
and finite range interactions preserve continuous symmetry \cite{N. D. Mermin PRL 1966}.
The quantum fluctuations, in general, originated from quantum effects,
are intrinsic and significant in low-dimensional magnetic systems
(LDMS). These are further prominent for low spin ($S=1/2$) magnetic
materials. The physics of $S=1/2$ LDMS is quite rich, and they offer
a viable ground for the experimental realization of correlated quantum
states with exotic fractional excitations \cite{A.Vasiliev NPJ 2018,A. Vasiliev handbook 2019}.
The algebraic spin-spin correlation decay in $S=1/2$ uniform spin
chain systems suggests that the ground state is gapless \cite{D. C. Johnston PRB 2000,J. Schlappa Nature 2012,M. Mourigal Nature 2013}.
Further, the introduction of geometric frustration through the presence
of $2^{nd}$ nearest neighbor (NN)\textbf{ }interaction ($J_{2}$)
along with that of $1^{st}$ NN interaction ($J_{1}$) to the $S=1/2$
uniform spin chain leads to a gapped excitation spectrum in the ground
state as per the exactly solvable Majumdar-Ghosh (MG) chain model
\cite{C.K. Majumdar JMP 1969}. The MG chain model states that $J_{2}/J_{1}=0.5$
opens a spin-gap and forms a singlet ground state \cite{S. Lebernegg PRB 2017}\textbf{. }

On the other hand, the LDMS with large spin (i.e., $S=5/2$) has not
been studied extensively as the quantum effects are not prominent
in these materials. A few examples with $S=5/2$ systems are SrMn$_{2}$V$_{2}$O$_{8}$
\cite{A.K. Bera PRB 2014}, SrMn(VO$_{4}$)(OH) \cite{L. D. Sanjeewa PRB 2016},
Ba$_{3}$Fe$_{2}$Ge$_{4}$O$_{14}$ \cite{L. D. Sanjeewa JSSC 2020},
and Bi$_{2}$Fe(SeO$_{3}$)OCl$_{3}$ \cite{P. S. Berdonosov Inorg. Chem  2014}.
For example, the Mn-based linear chain system SrMn$_{2}$V$_{2}$O$_{8}$
exhibits a broad maximum ($T^{max}$) in the susceptibility data $\chi(T)$
around $200$ K. However, due to the presence inter-chain couplings
($J'/J_{1}$$\geq0.6$), this material shows a magnetic long-range
order (LRO) at $45$ K. The Fe-based zig-zag chain Bi$_{2}$Fe(SeO$_{3}$)OCl$_{3}$
system shows $T^{max}$ around $130$ K in the $\chi(T)$ data and
LRO at $13$ K, even in the presence of magnetic frustration with
$J_{2}/J_{1}\thickapprox0.2$. All these $S=5/2$ spin systems undergo
LRO at finite temperature due to the inter-chain coupling and insufficient
magnetic frustration. It is pertinent to ask whether a correlated
dynamic ground state is realizable or not in low dimensional systems
with large spin. In this context, exploring novel low dimensional
spin systems with large spin $S=5/2$ promising to host quantum spin
liquid with exotic fractional excitations set an attractive setting.

In this paper, we report the magnetic properties and electronic structure
calculations on Bi$_{3}$FeMo$_{2}$O$_{12}$ \cite{A.W. Sleight materia research 1974}.
This material comprises the very well separated $S=5/2$ zig-zag chains
passing along the $c$-axis. The magnetic moments ($S=5/2$) interact
antiferromagnetically with Curie-Weiss temperature $\theta_{CW}\approx-40$
K. No magnetic LRO or spin freezing is observed down to $0.2$ K.
From the electronic structure calculations, the estimated ratio of
$2^{nd}$ NN and $1^{st}$ NN couplings between Fe atoms is close
to $1.1$, and negligible inter-chain coupling ($J'/J_{1}\approx0.01$),
which suggests that the Bi$_{3}$FeMo$_{2}$O$_{12}$ is a unique
material with very well separated $S=5/2$ triangular chains with
a small anisotropy. Interestingly, heat capacity shows a nearly linear
behavior with a finite value of the linear coefficient, suggesting
the gapless excitations in the $S=5/2$ triangular chains, unlike
the $S=1/2$ triangular chains that host a spin-gap ground state \cite{Uematsu 2020}.

\section{Experimental details}

The polycrystalline samples of Bi$_{3}$FeMo$_{2}$O$_{12}$ and the
non-magnetic analog Bi$_{3}$GaMo$_{2}$O$_{12}$ were synthesized
by solid-state reaction method using the respective chemicals of Bi$_{2}$O$_{3}$,
Fe$_{2}$O$_{3}$, Ga$_{2}$O$_{3}$, and MoO$_{3}$. These chemicals
are mixed in stoichiometric ratio and thoroughly grounded in agate
mortar and pestle. The pellets were made using the hydraulic press
and heated at different temperatures from $400$°C to $800$°C. Finally,
the sample was fired at $800$°C for $72$ hours with a few intermediate
grindings to obtain the single phase of the samples Bi$_{3}$FeMo$_{2}$O$_{12}$.
Neutron powder diffraction (NPD) experiments were performed using
the neutron powder diffractometer PD-I ( \textgreek{l}= $1.094$ Å
) with three linear position-sensitive detectors at Dhruva reactor,
Bhabha Atomic Research Center, India. Magnetization ($M$), and heat
capacity ($C_{p}$) measurements were performed on the polycrystalline
sample pellets using the Physical Properties Measurement System (PPMS)
with the corresponding attachments of Vibration Sample Magnetometer
(VSM) and heat-capacity measurement option, respectively, in the temperature
range from $2$ K to $300$ K and in the magnetic fields up to $160$
kOe. Low-temperature heat capacity data was measured using a dilution
fridge on a flat pellet of Bi$_{3}$FeMo$_{2}$O$_{12}$.

\section{Results}

\textbf{A. Structural details }

\begin{figure}
\includegraphics[scale=0.06]{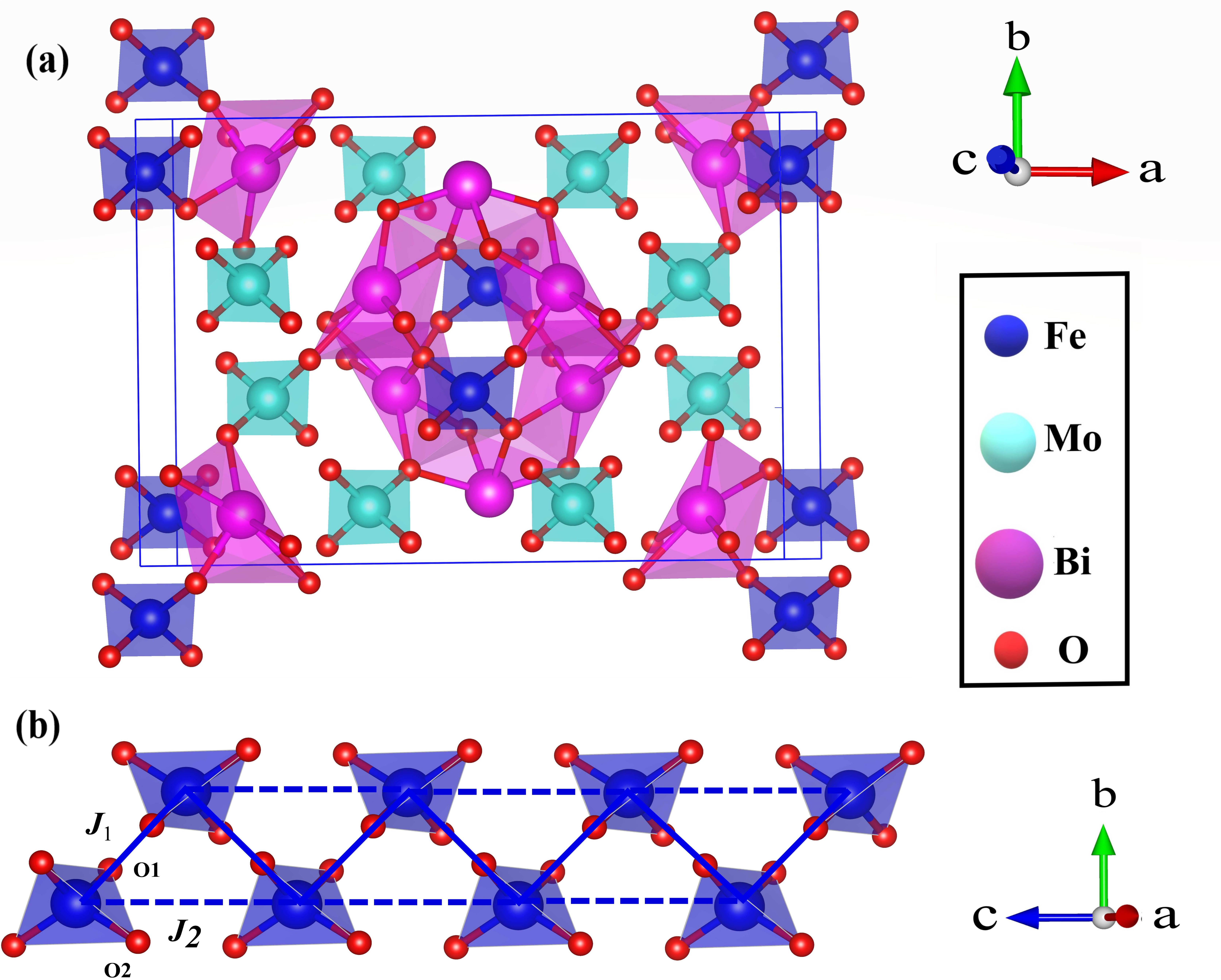}

\caption{\textcolor{black}{(Color online)}\label{fig1_CRYST}\textcolor{black}{{}
(a) Unit cell of Bi$_{3}$FeMo$_{2}$O$_{12}$} \cite{A.W. Sleight materia research 1974}.
(b) FeO$_{4}$ tetrahedral units form the zig-zag chain running along
$c$-axis. The 1$^{st}$ NN Fe-Fe distance is $3.8$ Å (with exchange
coupling $J_{1}$) and 2$^{nd}$ NN Fe-Fe distance is $5.3$ Å(with
exchange coupling $J_{2}$).}
\end{figure}

Bi$_{3}$FeMo$_{2}$O$_{12}$ crystallizes in a monoclinic structure
with a space group $C2/c$ and holds the Scheelite-type structure
with ABO$_{4}$ family \cite{A.W. Sleight materia research 1974}.
The Bi atoms are located at the A-site, while B-site is occupied by
Fe and Mo atoms in Bi$_{3}$FeMo$_{2}$O$_{12}$ structure. Interestingly,
the Fe and Mo atoms are ordered at the B-site. The unit cell consists
of FeO$_{4}$, MoO$_{4}$ tetrahedral, and BiO$_{6}$ polyhedral units
(see Fig. \ref{fig1_CRYST}($a$)).\textbf{ }The obtained lattice
parameters from the Rietveld refinement of XRD pattern are $a=16.91$
Å, $b=11.65$ Å, and $c=5.25$ Å, $\alpha=\gamma=$ $90$$\lyxmathsym{\textdegree}$,
$\beta$$=$ $107.1$$\lyxmathsym{\textdegree}$.The Fe$^{3+}$ $(S=5/2$)
ions form an infinite zig-zag chain running along the $c$-axis, as
shown in Fig. \ref{fig1_CRYST}($b$). The $1^{st}$ NN distance of
Fe-Fe is $3.79$ Å with a possible exchange path of Fe-O1-O1-Fe. The
$2^{nd}$ NN distance of Fe-Fe is $5.25$ Å, and its possible exchange
path could be through Fe-O2-O2-Fe interactions. The bond lengths and
bond angles are shown in the table \ref{exchange coupling}. These
chains are very well separated by a relatively large distance of $8.64$
Å, suggesting that the compound might have nearly isolated $S=5/2$
zig-zag spin chains.

\textbf{B. Neutron diffraction measurements}

\begin{figure}
\includegraphics[scale=0.33]{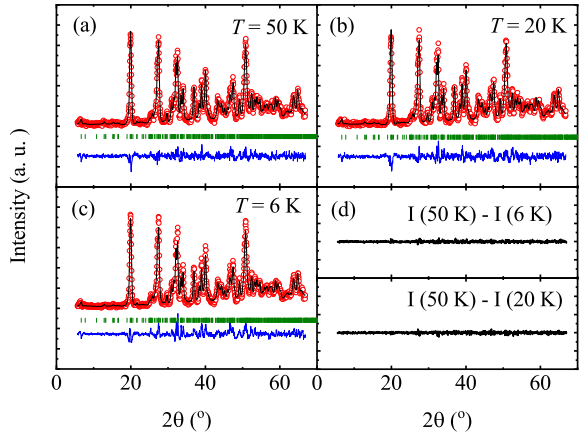}

\caption{(Color online) \label{ND data} Rietveld refinement of Neutron diffraction
(ND) data measured at different temperatures $50$ K, $20$ K, and
$6$ K are shown in ($a$), ($b$), and ($c$), respectively. Red
circles represent the measured data ($I_{obs}$), the black line represents
the calculated diffraction pattern ($I_{cal}$), the blue line represents
the difference ($I_{obs}$-$I_{cal}$), and the green vertical lines
represent the Bragg positions. (d) The ND intensities of $6$ K and
$20$ K after subtracting the ND intensities at $50$ K.}
\end{figure}

The data were recorded at different temperatures from $6$ K to $50$
K. Rietveld refinement of NPD data was performed using the FullProf
Suite software package as shown in Fig. \ref{ND data}($a$), ($b$),
and ($c$). The observed neutron diffraction pattern could be fitted
by considering only the nuclear phase. We have subtracted the intensities
of $50$ K from $6$ K data, i.e., $I$ ($50$ K) \textendash{} $I$
($6$ K). We do not see any signs of new Bragg intensities or diffuse
background patterns from the subtracted data as shown in Fig. \ref{ND data}($d$).
We have also compared the difference plot of $I$ ($50$ K) \textendash{}
$I$ ($6$ K) with the difference plot of $I$ ($50$ K) \textendash{}
$I$ ($20$ K), and there is no difference seen between these two
plots. Neither additional magnetic Bragg peaks nor an enhancement
in the intensity of the fundamental nuclear Bragg peaks has been observed
down to $6$ K (see Fig. \ref{ND data}($d$)), indicating the absence
of magnetic long-range order and rule out the presence of a phase
transition at $11$ K as was reported in reference \cite{C.Li  Cry 2019}. 

\textbf{C. Magnetization measurements}

\begin{figure}
\includegraphics[scale=0.45]{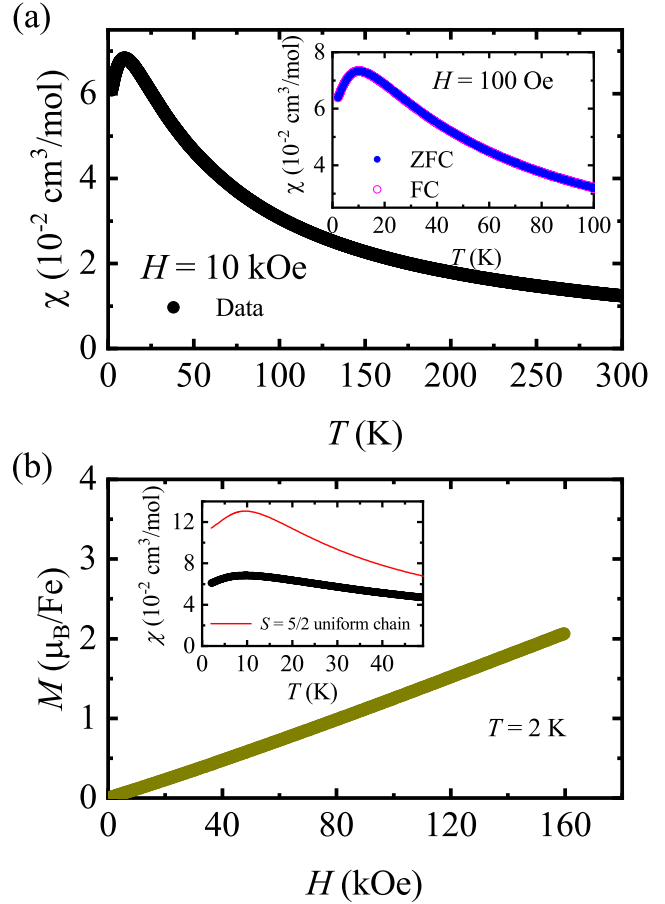}\caption{(Color online) \label{fig_chi and MH} (a) Temperature-dependent magnetic
susceptibility from $2$ K to $300$ K. The inset of figure (a) shows
the ZFC and FC magnetic susceptibility data under $100$ Oe field.
(b) Magnetic isotherm measured at $T$ = $2$ K with variation of
field $160$ kOe. The inset shows the comparison of experimental data
with $S=5/2$ uniform spin chain simulation for $J/k_{B}\approx-1$
K.}
\end{figure}

Temperature-dependent magnetic susceptibility $\chi(T)$ measurements
were performed on the polycrystalline sample Bi$_{3}$FeMo$_{2}$O$_{12}$
in the $T$ range from $2$ - $300$ K in $H=10$ kOe (see Fig. \ref{fig_chi and MH}($a$)).
The fit of the $\chi(T)$ data to Curie-Weiss law yields the temperature-independent
susceptibility $\chi_{0}\,\approx-2.0\times10^{-4}$ cm$^{3}$/mol,
Curie-Weiss temperature $\theta_{CW}\approx-\,40$ K, and Curie constant
$C\approx4.3$ cm$^{3}$ K/mol. Diamagnetic susceptibility is estimated
to be $\chi_{dia}\,\approx\,2.4\times10^{-4}$ cm$^{3}$/mol from
the individual ions in the formula Bi$_{3}$FeMo$_{2}$O$_{12}$.
After the subtraction of $\chi_{dia}$ from the obtained value of
$\chi_{0}$, the calculated Van Vleck susceptibility is $\chi_{vv}\,\approx\,4.2\times10^{-5}$cm$^{3}$/mol.
From the value of $C$, the effective magnetic moment of Fe$^{3+}$
is calculated to be $5.90$ $\lyxmathsym{\textmu}_{B}$ (= $\sqrt{8C}$
$\lyxmathsym{\textmu}_{B}$), well consistent with the expected value
of $5.91$ $\lyxmathsym{\textmu}$$_{B}$ for $S=5/2$. The absence
of splitting in zero-field-cooled (ZFC) and field-cooled (FC) susceptibility
rules out the spin-glass transition in this compound, as shown in
the inset of Fig. \ref{fig_chi and MH}($a$). Fig. \ref{fig_chi and MH}($b$)
represents the magnetization isotherm at $2$ K was measured up to
$160$ kOe. The $M(H)$ follows linear behavior, indicating the absence
of ferromagnetic components in the samples. $M(H)$ data do not saturate
up to $160$ kOe field. The large magnetic field is required to reach
saturated magnetization $M_{sat}=gS=5\mu_{B}$ for $S=5/2$ magnetic
monents.

At low-$T$, $\chi(T)$ shows a broad maximum around \textbf{$T^{max}\boldsymbol{\approx}10$}
K, indicating the presence of short-range correlations \cite{D. C. Johnston PRB 2000,R.nath PRB 2005,B. Koteswararao PRB2014,Y.Okamoto PRL 2007}.
We tried to analyze $\chi(T)$ with $S=5/2$ uniform spin chain model
by Bonner and Fisher (BF) \cite{M.E. Fisher Phy.Rev 1964,M. F. Fisher-1964}
and found that this model could not reproduce our experimental data.
According to the $S=5/2$ uniform spin chain, the broad maximum would
appear at $k_{B}$$T^{max}$/$J$ =$10.6$ \cite{L. J. De Jongh 2001 Adva. in Phy}.
From our experimental observation of $T^{max}$ position, the $J/k_{B}$
value expected to be about $-1$ K \cite{L. J. De Jongh 2001 Adva. in Phy}.
We have then compared the experimental data with the $S=5/2$ uniform
chain model with $J/k_{B}\thickapprox-1$ K. As shown in the inset
of Fig. \ref{fig_chi and MH}($b$), the experimental value of magnetic
susceptibility is much smaller than the simulated data, suggesting
the presence of significant additional antiferromagnetic exchange
couplings.\textbf{ }From this analysis, we anticipated the presence
of a significant value of $2^{nd}$ NN coupling along with $1^{st}$
NN coupling qualitatively. This is further quantitatively confirmed
from the first-principle density functional theory (DFT) electronic
structure calculations discussed later in the paper. The presence
of $2^{nd}$ NN exchange coupling accounts for the large magnetic
frustration in this material since $J_{1}$, and $J_{2}$ form a triangular
network (see Fig. \ref{fig1_CRYST}($b$)).

\textbf{D. Heat capacity measurments}

\begin{figure}
\includegraphics[scale=0.35]{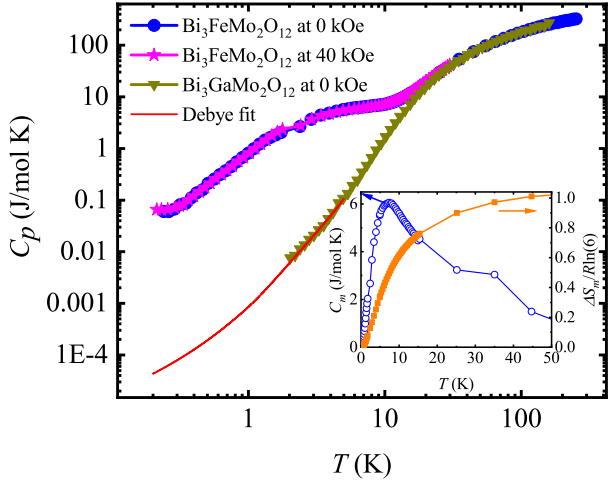}

\caption{(Color online)\label{fig:4 Cp HTdata}Temperature-dependent heat capacity
$C_{p}(T)$ data of Bi$_{3}$FeMo$_{2}$O$_{12}$ and Bi$_{3}$GaMo$_{2}$O$_{12}$
with lattice part of the heat capacity data (red) extracted from Debye
fit. The observed lattice contribution of the non-magnetic analog
is extremely small at low temperatures, comprated to the heat capacity
of Bi$_{3}$FeMo$_{2}$O$_{12}$. Inset shows the magnetic heat capacity
$C_{m}$ (left) and normalized magnetic entropy $\Delta S_{m}$( right)
versus $T$.}
\end{figure}

\begin{figure}
\includegraphics[scale=0.4]{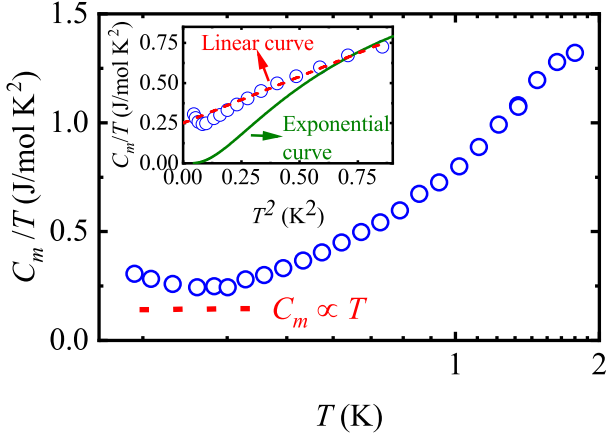}\caption{(Color online)\label{Cp lowTdata} $C_{m}/T$ vs. $T$ plot. Inset
shows $C_{m}/T$ vs. $T^{2}$ with the comparision of linear and exponential
curves.}
\end{figure}

The heat capacity $C_{p}(T)$ data of Bi$_{3}$FeMo$_{2}$O$_{12}$
and Bi$_{3}$GaMo$_{2}$O$_{12}$ were investigated in zero field
(see Fig. \ref{fig:4 Cp HTdata}). At low-$T$, there is a large difference
seen between the $C_{p}$ of Bi$_{3}$FeMo$_{2}$O$_{12}$ and Bi$_{3}$GaMo$_{2}$O$_{12}$
indicaitng the dominance of magnetic contribution. Interestingly,
no sharp peak is observed down to $0.2$ K in the $C_{p}(T)$ data
of Bi$_{3}$FeMo$_{2}$O$_{12}$, suggesting the absence of magnetic
LRO. The frustration parameter $f$ = $\mid\theta_{CW}\mid/T_{N}$
value is greater than $200$, indicating the presence of strong spin
frustration. To extract the magnetic contribution $C_{m}(T)$, we
have used the $C_{p}$ data of its non-magnetic analog Bi$_{3}$GaMo$_{2}$O$_{12}$
\cite{lattice subtraction reg}. $C_{p}$ data of Bi$_{3}$GaMo$_{2}$O$_{12}$
is fitted with Debye expression \cite{C. Kittle book} below $5$
K.

\textbf{$C_{ph}(T)=9R\sum c_{n}\left(\frac{T}{\theta_{Dn}}\right)^{3}\intop\frac{x^{4}e^{^{x}}}{(e^{x}-1)^{2}}dx$}

Here, the $\theta$$_{Dn}$ represent the Debye temperatures, $c_{n}$
indicate the multiplication coefficients, and $R$ is the universal
gas constant. The extracted Debye temperatures are $\theta$$_{D1}\approx\,$$240$
K and $\theta$$_{D2}\approx\,$$308$ K. The fit is extrapolated
down to $0.2$ K, and then subtracted from the $C_{p}$ data of Bi$_{3}$FeMo$_{2}$O$_{12}$.
As shown in the inset of Fig. \ref{fig:4 Cp HTdata}, the $C_{m}(T)$
data show a broad maximum at $6$ K, like at $10$ K in the $\chi(T)$
data. The appearance of a broad maximum in the $C_{p}$ data is due
to the short-range spin correlations that arise from the one-dimensional
nature of exchange interaction between Fe$^{3+}$ moments \cite{D. C. Johnston PRB 2000,R.nath PRB 2005}.
It has been noticed in many LDMS that the broad maximum in the heat
capacity is generally at low temperatures than that of the broad maximum
in susceptibility \cite{D. C. Johnston PRB 2000,B. Koteswararao JPCM 2015}.
The magnetic entropy $S_{m}$ is calculated from the integration of
$C_{m}/T$ versus $T$. The estimated entropy $S_{m}$ increases and
saturates to the maximum value of about $14.83$ J/mol K ($=R$ ln
$6$), expected for an $S=5/2$ system. 

The inset of Fig. \ref{Cp lowTdata} shows the plot of $C_{m}/T$
versus $T^{2}$. We have compared the data with the linear and exponential
curves. The data do not follow the exponential behavior, ruling out
the existence of a spin-gap in the ground state. A very small upturn
in $C_{m}/T$ versus $T^{2}$ plot indicates that the system might
be reaching a static magnetic long-range ordered state at extremely
low temperatures. In Fig. \ref{Cp lowTdata}, The $C_{m}/T$ is nearly
independent of $T$ at low temperatures($T$$\ll J/k_{B}$), indicating
that data follows nearly linear behavior expected for the systems
with gapless excitations. From the $C_{m}/T$ versus $T^{2}$ plot,
the intercept value is found to be $\gamma\approx250$ mJ/mol-K$^{2}$.
It is somewhat larger than that of $S=5/2$ linear chain system tetramethyl
ammonium manganese trichloride (TMMC) with $J/k_{B}$= $-6.7$ K (
$\gamma\approx9.8$ mJ/mol-K$^{2}$) \cite{W. J.M. de jonge prb 1975}.
From the normalized magnetic heat capacity ($C_{m}J/Nk_{B}^{2}$)
\cite{D. C. Johnston PRB 2000}, the scaled $\gamma$ value ($i.e.$
$\gamma J/Nk_{B}^{2}$) for Bi$_{3}$FeMo$_{2}$O$_{12}$ ($S=5/2$
triangular chain) is different from TMMC ($S=5/2$ linear chain).
The results suggest that $S=5/2$ triangular spin chain system Bi$_{3}$FeMo$_{2}$O$_{12}$
hosts the robust gapless excitations.

\textbf{E. DFT calculations}

To shed light on the electronic structures and to understand the magnetic
behavior of Bi$_{3}$FeMo$_{2}$O$_{12}$, spin-polarized DFT calculations
in the local-spin density approximation (LSDA) and LSDA+$U$ (Hubbard
$U$) approach were carried out by means of a full-potential linearized
muffin-tin orbital (FP-LMTO) method \cite{O. K. Andersen PRB 1974,J. M. Wills PRB 1987}
as implemented in the RSPt code \cite{J. M. wills Springers 2000}.
We have considered on-site Coulomb interaction $U$ = $2$ eV combined
with Hund's exchange $J_{H}$ = $0.8$ eV within the fully rotationally
invariant LSDA+$U$ approach \cite{V. I.Anisimov PRB 1993} to treat
the electronic correlation effects of Fe-$d$ states. Such choices
of $U$ are guided by a previous report on a correlated oxide containing
high-spin Fe$^{3+}$ions \cite{J. Chakraborty JMMM 2019}. From both
LSDA and LSDA+ $U$ calculations, the lowest energy state is identified
for the pattern with the antiferromagnetic couplings between the $1^{st}$
NN and $2^{nd}$ NN Fe-spins in the isolated zig-zag-chain.

\begin{figure}
\includegraphics[scale=0.32]{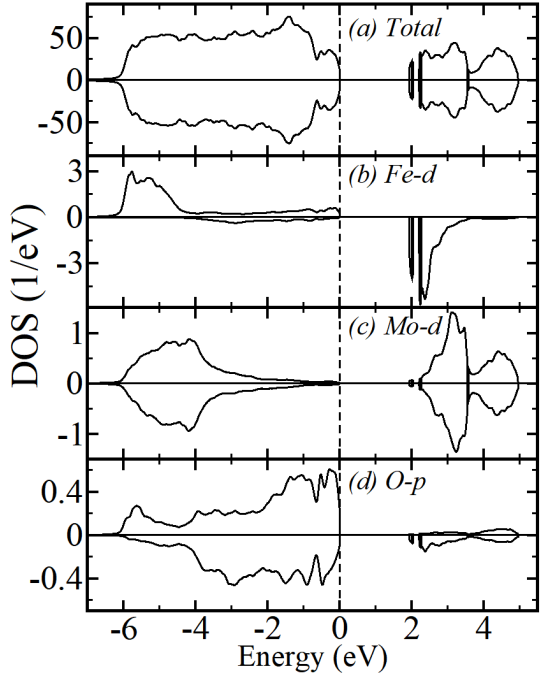}\caption{\label{DOS figure}Total and orbital-decomposed density of states
(DOS) for the lowest energy magnetic state. Fermi energy is set to
zero. }
\end{figure}

The computed total and orbital-decomposed density of states (DOS)
in this lowest-energy magnetic state as obtained from LSDA+$U$ are
shown in Fig\textbf{ }\ref{DOS figure}(a). The DOS of Fe-3$d$ clearly
shows that the majority spin states are fully filled up, and the minority
spin states are empty (see Fig.\ref{DOS figure}(b)). This is consistent
with the picture of the high-spin ground state of Fe$^{3+}$ ions
(3$d^{5}$). The Mo-$d$ states (see Fig. \ref{DOS figure}(c)) are
found to be completely empty, indicating the nominal $d^{0}$ non-magnetic
state of these ions. The O-$p$ states are also delocalized and spread
in the entire valence band. The insulating gap of $1.7$ eV is found
between O-$p$ and the Fe-$d$ states, making this system a member
of the charge-transfer insulator in the Zaanen, Sawatzky, and Allen
(ZSA) scheme \cite{J. Zaanen PRL 1985,S. Nimkar PRB 1993}. 

\begin{table}
\caption{\label{exchange coupling} The details of exchange coupling strengths
using LSDA approach with $U$}
\begin{tabular}{|>{\centering}p{2cm}|>{\centering}p{3.2cm}|>{\centering}p{3.2cm}|}
\hline 
 & $J_{1}$ & $J_{2}$\tabularnewline
\hline 
\hline 
Fe-Fe distance & $3.8$$\textrm{Å}$ & $5.3$$\textrm{Å}$\tabularnewline
\hline 
Exchange path & Fe-O1-O1-Fe

(Fe-O1-O1=$128.5$$\lyxmathsym{\textdegree}$\&

O1-O1-Fe=$71.9$$\lyxmathsym{\textdegree}$) & Fe-O2-O2-Fe

(Fe-O2-O2=$111.8$$\lyxmathsym{\textdegree}$\&

O2-O2-Fe=$111.8$$\lyxmathsym{\textdegree}$)\tabularnewline
\hline 
LSDA+U & -$1.52$ K & -$1.66$ K\tabularnewline
\hline 
\end{tabular}
\end{table}

We next estimated the various magnetic exchange couplings ($J_{1}$
and $J_{2}$ as marked in Fig. \ref{fig1_CRYST}) based on the converged
lowest energy magnetic state. Here we employed the formalism of Ref.
\cite{Y. O. Kvashnin PRB 2015}, where the total converged energies
of the magnetic system are mapped onto a Heisenberg Hamiltonian, and
magnetic force theorem \cite{A. Liechtenstein JMMM 1987,M. I. Katsnelson PRB 2000}
is applied to extract the inter-atomic magnetic exchange interactions.
Our calculations, as summarized in table \ref{exchange coupling},
reveal that both $J_{1}$ and $J_{2}$ are antiferromagnetic. The
$J_{1}$ and $J_{2}$ exchange interactions are mediated through Fe-O-O-Fe
super-exchange paths with varied Fe-Fe distances ($3.79$ Å and $5.25$
Å), as mentioned in table\ref{exchange coupling}. Interestingly,
the magnitudes of both the exchanges come out to be almost equal ($J_{2}/J_{1}\thickapprox1.1$)
although the corresponding Fe-Fe distances are very different. We
also note that such a conclusion is very robust and independent of
the adopted methods and independent of $U$ choice within the LSDA+$U$
approach.

The highly delocalized nature of O-$p$ orbitals promotes such super-super
exchange interaction. Interestingly, the Fe-O-O angles that connect
the two $2^{nd}$ NN Fe are equal while these angles are different
for $1^{st}$ NN. The peculiar character of the crystal symmetry is
responsible for the reason of having $J_{1}$ and $J_{2}$ to be nearly
equal, despite having a difference in the Fe-Fe bond distances. The
nature of the exchange could also be qualitatively understood within
the framework of the extended Kugel-Khomskii model \cite{K. I. Kugel ZETF 1973,K. I. Kugel SPU 1982}.
It is well established that half-filled orbitals promote antiferromagnetic
super-exchange since virtual hopping between Fe orbitals is allowed
only if they possess anti-parallel alignments. Importantly, the antiferromagnetic
nature, together with the comparable exchange interactions ($J_{1}$$\thickapprox$
$J_{2}$ ), causes strong spin-frustration in the zig-zag-shaped triangular
Fe$^{3+}$ chains. The estimated $\theta_{CW}\approx-37.5$ K agrees
very well with the experimental value of $-$$40$ K, providing further
credence to the theoretically computed values of the exchange interactions.
The ratio of the inter- to intra-chain magnetic exchange coupling
$J'/J_{1}$ is estimated at about $0.01$. Thus, we can conclude that
the present system is an example of a nearly isolated $S$ = $5/2$
triangular spin chain.

\section{Discussion}

Triangular antiferromagnets offer a promising ground for realizing
the unusual states of matter. In $S=5/2$ triangular systems, a few
2D magnets were investigated AFeO$_{2}$ (A=Cu, Li, and Na) \cite{F. Ye PRB 2006,M. Tabuchi Solid State Ionics 1995};
however, the quantum ground state without magnetic LRO has not been
identified. Achieving the disordered quantum state in $S=5/2$ systems
probably requires a material with ideal one-dimensionality. Our experimental
observations on $S=5/2$ triangular chain material Bi$_{3}$FeMo$_{2}$O$_{12}$
($J_{2}/J_{1}\thickapprox1.1$ and $J'/J\thickapprox0.01$) revealed
that it exhibits large magnetic frustration $(f>200$). Besides, the
observation of the linear behavior of $C_{m}(T)$ suggests the presence
of gapless excitations. Strikingly, the physics of $S=5/2$ triangular
chain behavior is entirely different from that of the $S=1/2$ triangular
chain, where one can expect a robust spin-gap ground state \cite{Uematsu 2020}.
As an example, $S=1/2$ zig-zag chain system Sr$_{0.9}$Ca$_{0.1}$CuO$_{2}$
has shown the spin-gap ground state due to the randomness \cite{Hammerath PRL 2011}.
All the results support that the titled compound might be a possible
candidate for gapless spin liquid. Muon spin relaxation and inelastic
neutron scattering experiments in sub-Kelvin temperature may shed
microscopic insights into the ground state and spin dynamics of the
titled material.

\section{Conclusion}

Our investigation reveals that the titled compound Bi$_{3}$FeMo$_{2}$O$_{12}$
is a nearly ideal quasi-one-dimensional $S=5/2$ triangular chain
system with a small anisotropy ($J_{2}/J_{1}$$\thickapprox$ $1.1$).
The presence of strong magnetic frustration and negligible inter-chain
interactions preclude magnetic LRO down to $200$ mK. $C_{m}(T)$
data show a linear behavior reflecting the gapless excitations in
the ground state. These results will stimulate both theoretical and
experimental interests to examine whether our $S=5/2$ triangular
chain Bi$_{3}$FeMo$_{2}$O$_{12}$ can host the spin liquid ground
state.

\textbf{\textit{Acknowledgments:}} B. K thanks DST INSPIRE faculty
award-2014 scheme. We thank Prof. P. L. Paulose and Dr. R. Kumar for
their support in magnetic measuemets. AKM thanks IIT Tirupati and
DST-SERB (Grant No.: ECR/2017/001903), Govt. of India, for providing
research grants. The work at SNU were supported by NRF(2019RIA2C2090648
and 2019M3E4A1080227). PK acknowledges the funding by the Science
and Engineering Research Board, and Department of Science and Technology,
India through Research Grants.


\begin{thebibliography}{10}
\bibitem{A.Vasiliev NPJ 2018}A. N. Vasiliev, O. Volkova, E. Zvereva,
and M. Markina, Milestones of low-D quantum magnetism, Npj Quantum
Mater. \textbf{3}, 1 (2018). 

\bibitem{A. Vasiliev handbook 2019}A. N. Vasilliev, O. S. Volkova,
E. A. Zvereva, and M. M. Markina, Low-Dimensional Magnetism, CRC Press-1st
ed., Handcover Book (2019).

\bibitem{L. Balents Nature 2010}L. Balents, Spin liquids in frustrated
magnets, Nature \textbf{464}, 199 (2010). 

\bibitem{C. Broholm Science 2020}C. Broholm, R. J. Cava, S. A. Kivelson,
D. G. Nocera, M. R. Norman, and T. Senthil, Quantum spin liquids,
Science \textbf{367}, 6475 (2020).

\bibitem{N. D. Mermin PRL 1966}N. D. Mermin and H. Wagner, Absence
of Ferromagnetism or Antiferromagnetism in One or Two-Dimensional
Isotropic Heisenberg Models, Phys. Rev. Lett. \textbf{17,} 1133 (1966).

\bibitem{D. C. Johnston PRB 2000}D. C. Johnston, R. K. Kremer, M.
Troyer, X. Wang, A. Klümper, S. L. Budko, A. F. Panchula, and P. C.
Canfield, Thermodynamics of spin $S$ = $1/2$ antiferromagnetic uniform
and alternating-exchange Heisenberg chains, Phys. Rev. B \textbf{61},
9558 (2000).

\bibitem{J. Schlappa Nature 2012}J. Schlappa, K. Wohlfeld, K. J.
Zhou, M. Mourigal, M. W. Haverkort, V. N. Strocov, L. Hozoi, C. Monney,
S. Nishimoto, S. Singh, and A. Revcolevschi, Spin\textendash orbital
separation in the quasi-one-dimensional Mott insulator Sr$_{2}$CuO$_{3}$,
Nature \textbf{485,} 82( 2012).

\bibitem{M. Mourigal Nature 2013}M. Mourigal, M. Enderle, A. Klöpperpieper,
J. S. Caux, A. Stunault, and H. W. Rønnow, Fractional spinon excitations
in the quantum Heisenberg antiferromagnetic chain, Nature Physics
\textbf{9}, 435 (2013).

\bibitem{C.K. Majumdar JMP 1969}C. K. Majumdar, and D. K. Ghosh,
On next-nearest-neighbor interaction in linear chain. II, J. Math.
Phys. \textbf{10,} 1399 (1969).

\bibitem{S. Lebernegg PRB 2017}S. Lebernegg, O. Janson, I. Rousochatzakis,
S. Nishimoto, H. Rosner, and A. A. Tsirlin, Frustrated spin chain
physics near the Majumdar-Ghosh point in szenicsite Cu$_{3}$(MoO$_{4}$)(OH)$_{4}$,
Phys. Rev. B \textbf{95,} 035145 (2017).

\bibitem{A.K. Bera PRB 2014}A. K. Bera, B. Lake, W. D. Stein, and
S. Zander, Magnetic correlations of the quasi-one-dimensional half-integer
spin-chain antiferromagnets SrM$_{2}$V$_{2}$O$_{8}$ (M = Co, Mn),
Phys. Rev. B \textbf{89}, 094402 (2014).

\bibitem{L. D. Sanjeewa PRB 2016}L. D. Sanjeewa, V. O. Garlea, M.
A. McGuire, C. D. McMillen, H. B. Cao, and J. W. Kolis, Structural
and Magnetic Characterization of the One-dimensional $S$ = $5/2$
Antiferromagnetic Chain System SrMn(VO$_{4}$)(OH), Phys. Rev. B \textbf{93},
224407 (2016). 

\bibitem{L. D. Sanjeewa JSSC 2020}L. D. Sanjeewa, A. S. Sefat, M.
Smart, M. A. McGuire, C. D. McMillen, J. W. Kolis, Synthesis, structure
and magnetic properties of Ba$_{3}$M$_{2}$Ge$_{4}$O$_{14}$(M=
Mn and Fe): Quasi-one-dimensional zig-zag chain compounds, Journal
of Solid State Chemistry \textbf{283, }121090 (2020).

\bibitem{P. S. Berdonosov Inorg. Chem  2014}P. S. Berdonosov, E.
S. Kuznetsova, V. A. Dolgikh, A. V. Sobolev, I. A. Presniakov, A.
V. Olenev, B. Rahaman, T. Saha-Dasgupta, K. V. Zakharov, E. A. Zvereva,
O. S. Volkova, and A. N. Vasiliev, Crystal structure, physical properties,
and electronic and magnetic structure of the spin $S=5/2$ zig-zag
chain compound Bi$_{2}$Fe(SeO$_{3}$)$_{2}$OCl$_{3}$, Inorganic
Chemistry \textbf{53}, 5830 (2014).

\bibitem{A.W. Sleight materia research 1974}A. W. Sleight, and W.
Jeitschko, Bi$_{3}$(FeO$_{4}$)(MoO$_{4}$)$_{2}$ and Bi$_{3}$(GaO$_{4}$)(MoO$_{4}$)$_{2}$-
new compounds with scheelite related structures, Materials Research
Bulletin \textbf{9,} 951 (1974).

\bibitem{Uematsu 2020}K. Uematsu, T. Hikihara, and H. Kawamura, Frustration-induced
quantum spin liquid behavior in the $S=1/2$ random-bond Heisenberg
antiferromagnet on the zig-zag chain, arXiv preprint arXiv:2009.08630
(2020).

\bibitem{C.Li  Cry 2019}C. Li, Z. Gao, X. Tian, J. Zhang, D. Ju,
Q. Wu, W. Lu, Y. Sun, D. Cui, and X. Tao, Bulk crystal growth and
characterization of the bismuth ferrite-based material Bi$_{3}$FeO$_{4}$(MoO$_{4}$)$_{2}$,
Cryst Eng Comm, \textbf{21}, 2508 (2019).

\bibitem{R.nath PRB 2005}R. Nath, A. V. Mahajan, N. Büttgen, C. Kegler,
A. Loidl, and J. Bobroff, tudy of one-dimensional nature of S = 1/2
(Sr, Ba)$_{2}$Cu(PO$_{4}$)$_{2}$ and BaCuP$_{2}$O$_{7}$ via $^{31}$P
NMR, Phys. Rev. B \textbf{71}, 174436 (2005).

\bibitem{B. Koteswararao PRB2014}B. Koteswararao, R. Kumar, P. Khuntia,
Sayantika Bhowal, S. K. Panda, M. R. Rahman, A. V. Mahajan, I. Dasgupta,
M. Baenitz, Kee Hoon Kim, and F. C. Chou, Magnetic properties and
heat capacity of the three-dimensional frustrated S = 1/2 antiferromagnet
PbCuTe$_{2}$O$_{6}$, Phys. Rev. B 90, 035141( 2014).

\bibitem{Y.Okamoto PRL 2007}Y. Okamoto, M. Nohara, H. Aruga-Katori,
and H. Takagi, Spin-Liquid State in the S = 1/2 Hyperkagome Antiferromagnet
Na$_{4}$Ir$_{3}$O$_{8}$, Phys. Rev. Lett. \textbf{99}, 137207 (2007).

\bibitem{M.E. Fisher Phy.Rev 1964}M. E. Fisher, and J. C. Bonner,
Linear Magnetic Chains with Anisotropic coupling, Phys. Rev. \textbf{135,}
A640 (1964). 

\bibitem{M. F. Fisher-1964}M. E. Fisher, Magnetism in one-dimensional
systems-the Heisenberg model for infinite spin, American Journal of
Physics \textbf{32}, 343 (1964).

\bibitem{L. J. De Jongh 2001 Adva. in Phy}L. J. De Jongh, and A.
R. Miedema, Experiments on simple magnetic model systems, Advances
in Physics \textbf{50}, 947 (2001).

\bibitem{lattice subtraction reg} To match both the $C_{p}$ data
at high temperatures from $200$ K to $250$ K, the $C_{p}$ data
of Bi$_{3}$GaMo$_{2}$O$_{12}$ is multplied with $1.09$ and then
used it for substracting the lattice of Bi$_{3}$FeMo$_{2}$O$\text{\ensuremath{_{12}}}$. 

\bibitem{C. Kittle book}C. Kittel, Introduction to Solid State Physics,
John Wiley and Sons press - 8$^{th}$ ed., New York, NY (2004).

\bibitem{B. Koteswararao JPCM 2015}B. Koteswararao, S. K. Panda,
R. Kumar, Kyongjun Yoo, A. V. Mahajan, I. Dasgupta, B. H. Chen, Kee
Hoon Kim, and F. C. Chou, Observation of $S$ = $1/2$ quasi-1D magnetic
and magneto-dielectric behavior in a cubic SrCuTe$_{2}$O$_{6}$,
J. Phys. Condens. Matter \textbf{27}, 426001 (2015).

\bibitem{W. J.M. de jonge prb 1975}W. J. M. de Jonge, C. H. W. Swüste,
K. Kopinga, and K. Takeda, Specific heat of nearly-one-dimensional
tetramethyl ammonium manganese trichloride (TMMC) and tetramethyl
ammonium cadmium trichloride (TMCC), Phys. Rev. B \textbf{12}, 5858
(1975).

\bibitem{O. K. Andersen PRB 1974}O. K. Andersen, Linear methods in
band theory, Phys. Rev. B \textbf{12}, 3060 (1975). 

\bibitem{J. M. Wills PRB 1987}J. M. Wills and B. R. Cooper, Synthesis
of band and model Hamiltonian theory for hybridizing cerium systems,
Phys. Rev. B \textbf{36}, 3809 (1987). 

\bibitem{J. M. wills Springers 2000}J. M. Wills, O. Eriksson, M.
Alouni, and D. L. Price, Electronic Structure and Physical Properties
of Solids: The Uses of the LMTO Method (Springer, Berlin, 2000). 

\bibitem{V. I.Anisimov PRB 1993}V. I. Anisimov, I. V. Solovyev, M.
A. Korotin, M. T. Czyzyk, and G. A. Sawatzky, Density-functional theory
and NiO photoemission spectra, Phys. Rev. B \textbf{48}, 16929 (1993). 

\bibitem{J. Chakraborty JMMM 2019}J. Chakraborty and I. Dasgupta,
First principles study of electronic structure, magnetism and ferroelectric
properties of rhombohedral AgFeO$_{2}$, J. Magn. Magn. Mater. \textbf{487},
165296 (2019). 

\bibitem{J. Zaanen PRL 1985}J. Zaanen, G. A. Sawatzky and J. W. Allen,
Band gaps and electronic structure of transition metal compounds,
Phys. Rev. Lett. \textbf{55}, 418 (1985). 

\bibitem{S. Nimkar PRB 1993}S. Nimkar, D. D. Sarma, H. R. Krishnamurthy,
and S. Ramasesha, Mean-field results of the multiple-band extended
Hubbard model for the square-planar CuO$_{2}$ lattice, Phys. Rev.
B \textbf{48}, 7355 (1993).

\bibitem{Y. O. Kvashnin PRB 2015}Y. O. Kvashnin, O. Granas, I. Di
Marco, M. I. Katsnelson, A. I. Lichtenstein, and O. Eriksson, Exchange
parameters of strongly correlated materials: Extraction from spin-polarized
density functional theory plus dynamical mean-field theory, Phys.
Rev. B \textbf{91}, 125133 (2015).

\bibitem{A. Liechtenstein JMMM 1987}A. Liechtenstein, M. Katsnelson,
V. Antropov, and V. Gubanov, Local spin density functional approach
to the theory of exchange interactions in ferromagnetic metals and
alloys, J. Magn. Magn. Mater. \textbf{67}, 65 (1987).

\bibitem{M. I. Katsnelson PRB 2000}M. I. Katsnelson and A. I. Lichtenstein,
First-principles calculations of magnetic interactions in correlated
systems, Phys. Rev. B \textbf{61}, 8906 (2000). 

\bibitem{K. I. Kugel ZETF 1973}K. I. Kugel and D. I. Khomskii, Crystal-structure
and magnetic properties of substances with orbital degeneracy, Zh.
Eksp. Teor. Fiz. \textbf{64}, 1429 (1973).

\bibitem{K. I. Kugel SPU 1982}K. I. Kugel and D. I. Khomskii, The
Jahn-Teller effect and magnetism: transition metal compounds, Sov.
Phys. Uspekhi \textbf{25}, 231 (1982).

\bibitem{F. Ye PRB 2006}F. Ye, Y. Ren, Q. Huang, J. A. Fernandez-Baca,
Pengcheng Dai, J. W. Lynn, and T. Kimura, Spontaneous spin-lattice
coupling in the geometrically frustrated triangular lattice antiferromagnet
CuFeO$_{2}$, Phys. Rev. B. \textbf{73}, 220404(R) (2006).

\bibitem{M. Tabuchi Solid State Ionics 1995}M. Tabuchi, K. Ado, H.
Sakaebe, C. Masquelier, H. Kageyama, and O. Nakamura, Preparation
of AFeO$_{2}$ (A= Li, Na) by hydrothermal method, Solid State Ionics
\textbf{79}, 220 (1995).

\bibitem{Hammerath PRL 2011}F. Hammerath, S. Nishimoto, H.-J. Grafe,
A. U. B. Wolter, V. Kataev, P. Ribeiro, C. Hess, S.-L. Drechsler,
and B. Büchner, Spin Gap in the zig zag Spin-1/2 Chain Cuprate Sr$_{0.9}$Ca$_{0.1}$CuO$_{2}$,
Phys. Rev. Lett. \textbf{107}, 017203 (2011).
\end{thebibliography}
\end{document}